\documentclass[usenatbib]{basi}
\usepackage[T1]{fontenc}
\usepackage[british]{babel}
\usepackage[varg]{txfonts}
\usepackage{graphicx,lscape, subfigure}
\usepackage{rotating}
\usepackage{dcolumn}
\begin{document}
\title[Equipartition magnetic fields]{Equipartition magnetic fields and star formation rates in normal galaxies at sub-kpc scales}
\author[A.~Basu \& S. Roy]%
       {Aritra Basu\thanks{email: \texttt{aritra@ncra.tifr.res.in}}
       and Subhashis Roy\\
       National Centre for Radio Astrophysics, Pune University Campus, Ganeshkhind Road, Pune-411007, India}

\pubyear{2014}
\volume{00}
\pagerange{\pageref{firstpage}--\pageref{lastpage}}

\date{To appear in proceedings of {\it ``The Metrewavelength Sky Conference''} held at 
NCRA-TIFR, Pune, India, from 9--13 December 2013.}

\maketitle
\label{firstpage}

\vspace{-5mm}
\begin{abstract}
We studied the total magnetic field strength in normal star-forming galaxies
estimated using energy equipartition assumption. Using the well known
radio--far infrared correlation we demonstrate that the equipartition
assumption is valid in galaxies at sub-kpc scales.  We find that the magnetic
field strength is strongly correlated with the surface star formation rate in
the galaxies NGC 6946 and NGC 5236.  Further, we compare the magnetic field
energy density to the total (thermal + turbulent) energy densities of gas
(neutral + ionized) to identify regions of efficient field amplification in the
galaxy NGC 6946.  We find that in regions of efficient star formation, the
magnetic field energy density is comparable to that of the total energy density
of various interstellar medium components and systematically dominates in
regions of low star formation efficiency.

\end{abstract}

\section{Introduction}\label{s:intro}
\vspace{-5mm}

It is believed that the magnetic field in star forming galaxies are amplified
and maintained by dynamo action. At small scales ($\lesssim1$ kpc), the
turbulent dynamo action, driven by Supernova shocks and star formation,
amplifies the magnetic field strength efficiently up to energy equipartition
values in $\sim 10^6 - 10^7$ years.  Such regions are characterized by
small-scale random field with low degree of polarization \citep[see
e.g.][]{brand12, arsha09, bhat13}.  At larger scales ($\sim1-10$ kpc), the mean
field dynamo action is responsible for amplification of the field in
$\sim10^8-10^9$ years giving rise to large-scale coherent field with high
degree of polarization \citep{arsha09, subra08, shuku06}.  Magnetic field
strength grows due to stretching and twisting of the field lines up to energy
equipartition values, i.e., energy density in magnetic field is similar to the
kinetic energy density due to turbulent motions of gas.  Numerical
magnetohydrodynamic simulations of turbulent interstellar medium (ISM) have
revealed that the magnetic field ($B$) and gas density ($\rho_{\rm gas}$) are
coupled as $B\propto\rho_{\rm gas}^\kappa$, where $\kappa$ is the coupling
index \citep{cho00, fiedl93, grove03}. Under the condition of equipartition of
energy, $\kappa$ assumes the value 0.5 \citep{cho00}. It is therefore expected
that the radial scale length of the magnetic field energy and that of gas
density should remain similar.  However, in contradiction, it is often observed
that the magnetic energy density dominates over the total energy density of gas
especially towards the outer parts of the galaxies \citep{basu13, beck07}.
Star formation activity and shocks arising from supernova explosions drives
turbulence at the smallest scale in the ISM.  It is therefore imperative to do
spatially resolved study of the various competing forces in the ISM with the
magnetic field strength based on star formation activity.

\section{Results}\label{result}
\vspace{-5mm}

The galaxies studied here were observed using the Giant Meterwave Radio
Telescope at 0.33 GHz \citep{basu12a} and archival data observed using the Very
Large Array at 1.4 GHz.  The nonthermal emission maps were derived by
subtracting the thermal free--free emission component at both the frequencies
using a new technique developed by Basu \& Roy (2013) at sub-kpc scales. It was
found that at 0.33 GHz more than 95 percent and at 1.4 GHz about 90 percent of
the total radio emission is nonthermal in origin. The nonthermal maps were used
to determine the nonthermal spectral index\footnote{Spectral index, $\alpha$,
is defined as, $S_\nu \propto \nu^{-\alpha}$. Here, $S_\nu$ is the flux density
at a radio frequency $\nu$.}, $\alpha_{\rm nt}$, for estimating the
equipartition magnetic field strength maps and the coupling index $\kappa$ from
the slope of the radio--far infrared (FIR) correlation. We used the revised
equipartition formula given by \citet{beck05} to estimate the total magnetic
field strength.  Further, to validate the energy equipartition assumption, we
study the spatially resolved radio--FIR correlation at sub-kpc scales for the
galaxy NGC 6946.

\vspace{-5mm}

\subsection{Radio--FIR correlation: A test for equipartition condition}
\vspace{-5mm}

Building on the model first proposed by \citet{nikla97}, \citet{dumas11} showed
that, the slope of the radio--FIR correlation (as determined using the
nonthermal part of the radio emission only), $b$, to be related with the
coupling index $\kappa$, through the Kennicutt-Schmidt (KS) law index, $n$
\citep{kenni98} and $\alpha_{\rm nt}$.  Depending of whether equipartition
conditions are valid or not, one expects linear or non-linear slope of the
radio--FIR correlation \citep{beck13, dumas11, basu12b}. 

\begin{figure}
\begin{center}
\begin{tabular}{p{5cm}cp{5cm}}
\raisebox{-\height}{\includegraphics[width=5.0cm]{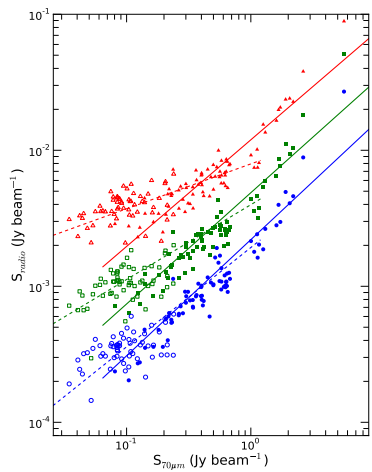}} & \quad &
\caption{Multi frequency radio--FIR correlation for NGC 6946. The red triangles,
green squares and blue circles are for 0.33, 1.4 and 5 GHz respectively.  The
filled and open symbols are for arm and interarm regions respectively.
\label{radiofir}}
\end{tabular}
\end{center}
\end{figure}

\citet{basu12b} from spatially resolved study at $\sim1$ kpc scales for four
galaxies (NGC 4736, NGC 5055, NGC 5236 and NGC 6946) found $\kappa=0.51\pm0.12$
and thereby concluded the equipartition conditions to be valid at scales of
$\sim1$ kpc. However, conversely, it could be presaged, if galaxies are in
energy equipartition, i.e., $\kappa \sim 0.5$, KS index $n=1.4$ \citep{kenni98}
and the nonthermal spectral index is constrained \citep[$\alpha_{\rm nt}\sim
0.8$;][]{basu12a}, then the slope of the radio--FIR correlation should be
$\sim0.8$. Moreover, the slope must remain the same when studied at different
radio frequencies, i.e., independent of the energy of the cosmic ray electrons
(CREs) giving rise to the correlation. This is indeed observed for the galaxy
NGC 6946 between 0.33 GHz and 5 GHz.  Fig.~\ref{radiofir} shows the radio
intensity at 0.33 GHz (red triangles), 1.4 GHz (green squares) and 5 GHz (blue
circles) vs. FIR intensity at $\lambda70~\mu$m. The open and filled symbols
represents for arm and interarm regions respectively. In the arm regions where
the star formation rate is higher, the slope is found to be $0.78\pm0.06$ at
0.33 GHz, $0.82\pm0.06$ at 1.4 GHz and $0.85\pm0.07$ at 5 GHz.  In the interarm
regions, i.e., regions of low star formation, the slope is observed to flatten
significantly at 0.33 and 1.4 GHz due to propagation of CREs from the regions
of generation in the arms to interarms \citep[see e.g.,][]{basu12b}.

\vspace{-5mm}
\subsection{Magnetic field and star formation activity}
\vspace{-5mm}

\begin{figure}
\begin{center}
\begin{tabular}{ccc}
\includegraphics[width=4.5cm]{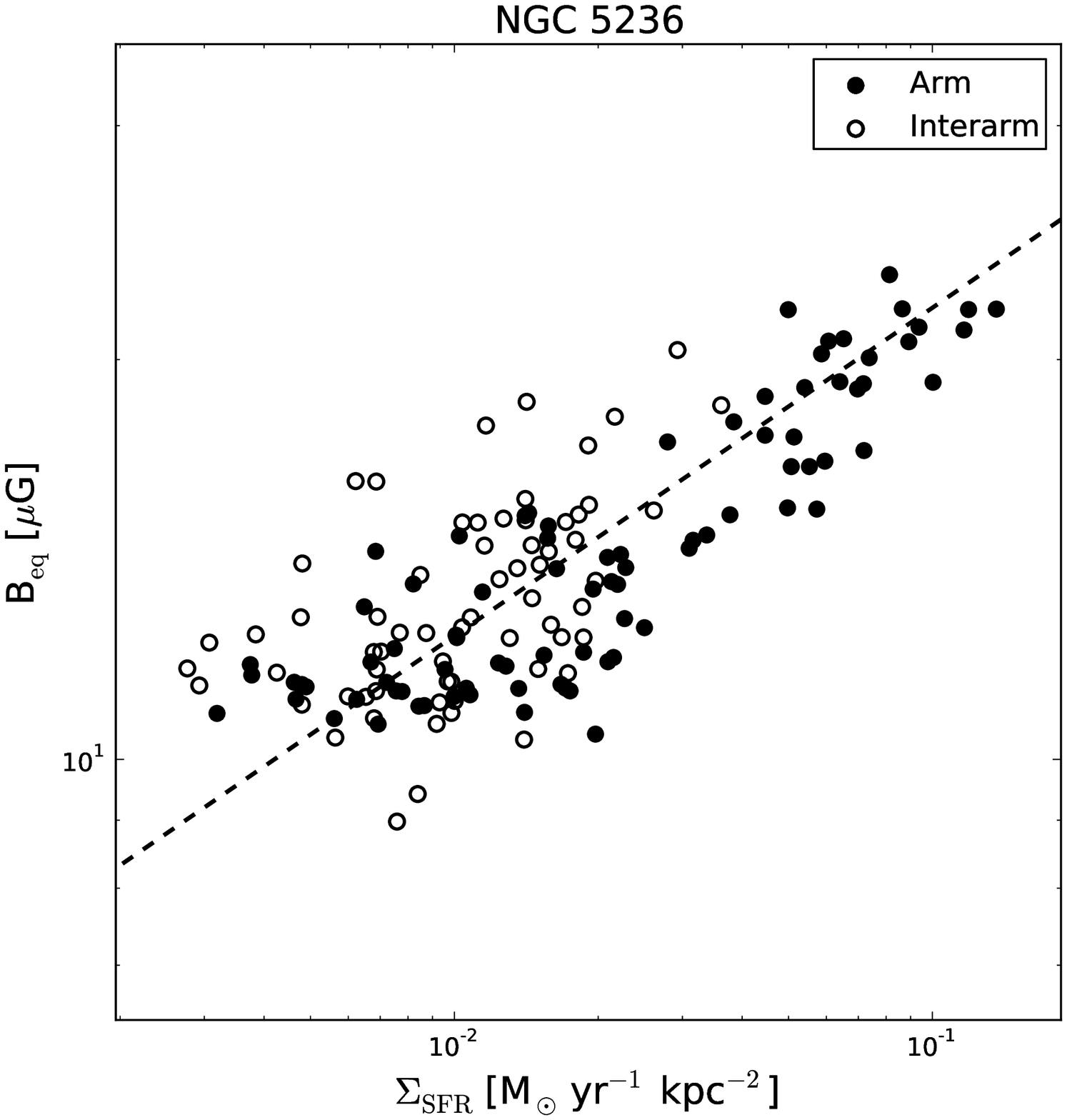} &
\includegraphics[width=4.5cm]{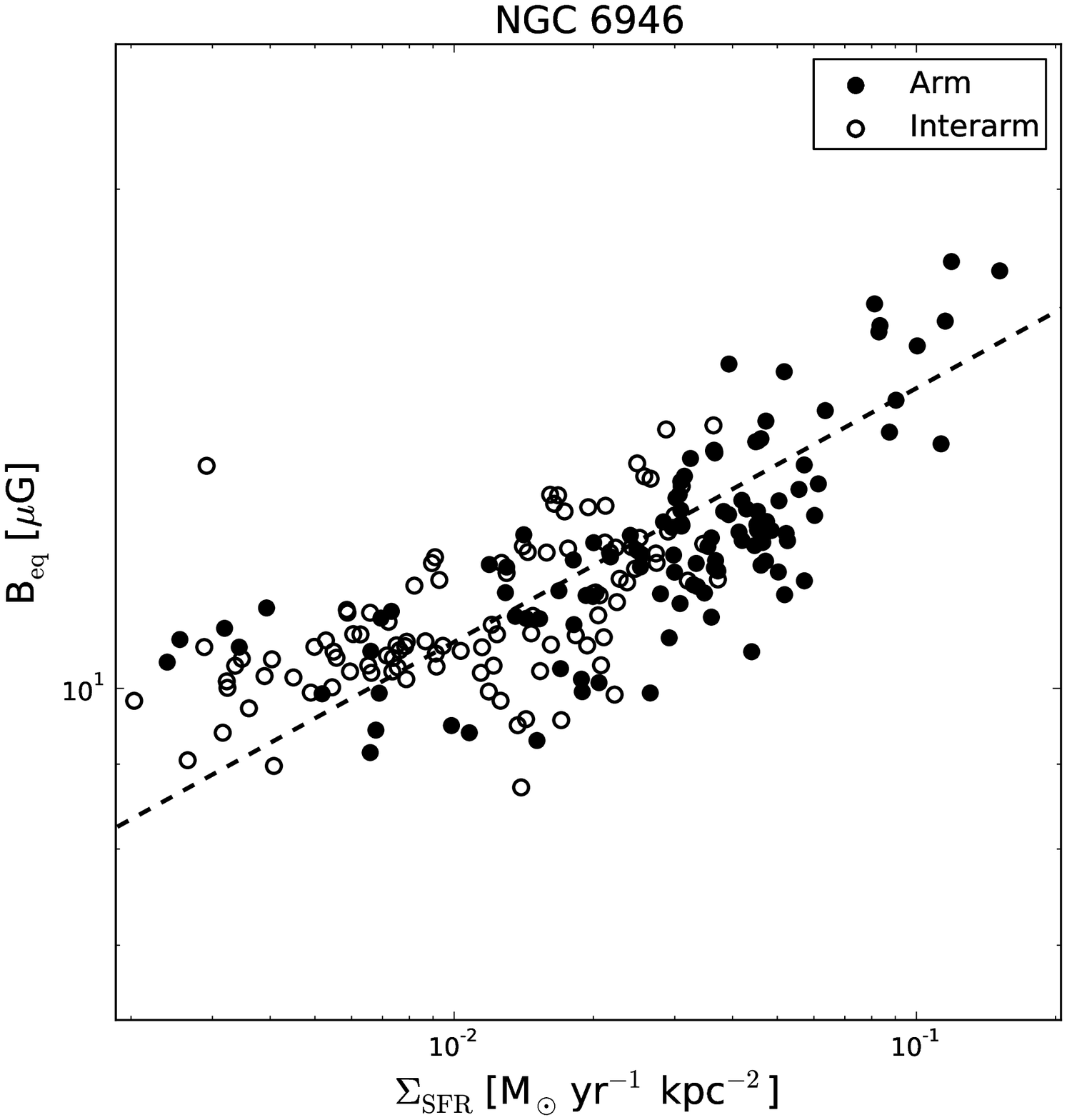} &
\includegraphics[width=4.5cm]{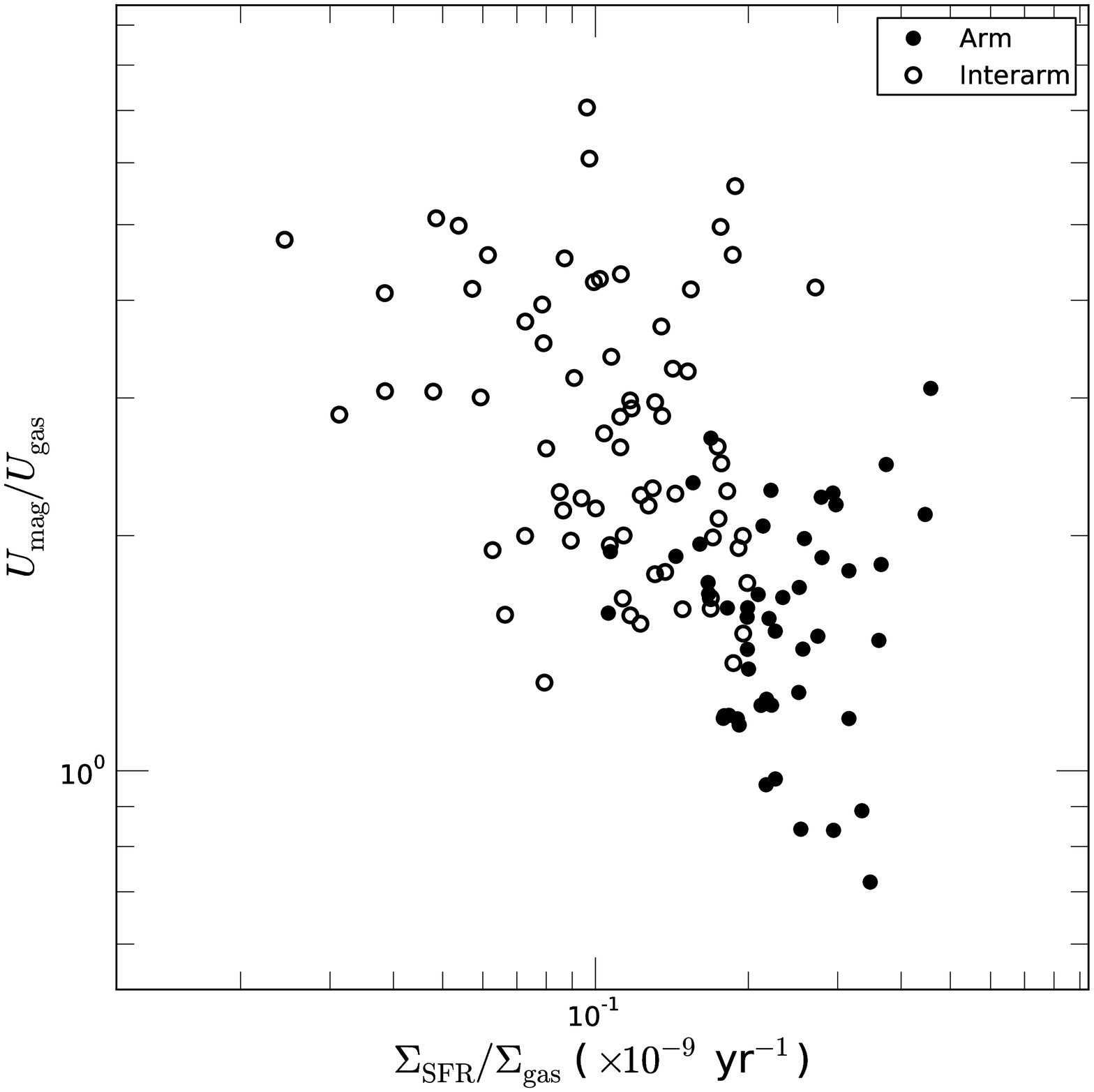}\\
(a) & (b) & (c) 
\end{tabular}
\caption{Variation of $B_{\rm eq}$ (in $\mu$G) with $\Sigma_{\rm SFR}$ (in $\rm
M_\odot yr^{-1} kpc^{-2}$) for the galaxy (a) NGC 6946 and (b) NGC 5236. Figure
(c) shows the variation of $U_{\rm mag}/U_{\rm gas}$ with $\Sigma_{\rm
SFR}/\Sigma_{\rm gas}$ (in $\rm yr^{-1}$) for the galaxy NGC 6946. The filled
and unfilled symbols are for arm and interarm regions
respectively.\label{beqsfr}} 
\end{center}
\end{figure}

Star formation activity in galaxies plays an important role in driving
turbulence and thereby amplification of magnetic fields.  Magnetic field
strength is expected to be correlated with the local star formation rate in
galaxies as supernova shocks amplify the field strength. In Fig.~\ref{beqsfr}
(a, b), we study the spatially resolved total magnetic field strength ($B_{\rm
eq}$) with the surface star formation rate ($\Sigma_{\rm SFR}$) at scales of
$\sim0.5$ kpc for the galaxies NGC 5236 (Fig. 2a) and NGC 6946 (Fig.  2b).  The
filled and unfilled circles are for the arm and interarm regions respectively.
Magnetic field is seen to be significantly correlated with the star formation
rate as a power law, i.e., $B_{\rm eq} \propto (\Sigma_{\rm SFR})^a$, where,
$a$ is the power law index. The Spearman's rank correlation for NGC 5236 is
0.80 and for NGC 6946 it is 0.74.  The power law index, $a$, is found to be
$0.25\pm0.05$ and $0.18\pm0.03$ for the galaxies NGC 5236 and NGC 6946
respectively. NGC 5236 shows comparatively steeper value of $a$ than NGC 6946,
indicating efficient magnetic field amplification, perhaps caused due to bar
action. However, more data are required to firmly establish our results.

In Fig. 2(c) we plot the ratio of the magnetic field energy density ($U_{\rm
mag} = B_{\rm eq}^2/8\pi$) to that of the total energy density of ISM gas
($U_{\rm gas}$) as a function of star formation efficiency ($\Sigma_{\rm
SFR}/\Sigma_{\rm gas}$) for the galaxy NGC 6946 at scales of $\sim0.5$ kpc.
Here, $\Sigma_{\rm gas}$ is the surface density of the total neutral gas.  The
total energy density of gas is computed from the thermal energy density of
neutral and ionized gas ($\frac{3}{2}\langle n\rangle kT$) and kinetic energy
due to turbulent motion of neutral (atomic+molecular) gas
\citep[$\frac{1}{2}\rho_{\rm gas}\delta v_{\rm turb}^2$; see][for
details]{basu13}. The filled and unfilled circles are for arm and interarm
regions respectively. It is seen that in the interarm regions, i.e., the
regions of low star formation efficiency, magnetic field energy density is
systematically higher than the total energy density of the various competing
forces in the ISM.  These are also the regions where large-scale coherent
magnetic field is observed \citep{beck07}. In the arm regions, i.e., regions of
high star formation efficiency, ISM is roughly in energy equipartition within a
factor of 2. Such regions are dominated by turbulence and perhaps the field is
amplified by small-scale turbulent dynamo action.

\end{document}